# Temporal Offsets between Maximum CME Speed Index and Solar, Geomagnetic, and Interplanetary Indicators during Solar Cycle 23 and the Ascending Phase of Cycle 24

A. Özgüç[1], A. Kilcik[2], K. Georgieva[3], B. Kirov[3]

[1] Kandilli Observatory and Earthquake Research Institute, Bogazici University, Istanbul, Turkey

[2] Faculty of Science, Department of Space Science and Technologies, Akdeniz University, Antalya, Turkey

[3] Space Research and Technology Institute, Bulgarian Academy of Sciences, Sofia, Bulgaria

Abstract: On the basis of morphological analysis of yearly values of the maximum CME (coronal mass ejection) speed index, the sunspot number and total sunspot area, sunspot magnetic field, and solar flare index, the solar wind speed and interplanetary magnetic field strength, and the geomagnetic $A_p$ and $D_{st}$ indices, we point out the particularities of solar and geomagnetic activity during the last cycle 23, the long minimum which followed it and the ascending branch of cycle 24. We also analyze temporal offset between the maximum CME speed index and the above-mentioned solar, geomagnetic, and interplanetary indices. It is found that this solar activity index, analyzed jointly with other solar activity, interplanetary parameters, and geomagnetic activity indices, shows a hysteresis phenomenon. It is observed that these parameters follow different paths for the ascending and the descending phases of solar cycle 23. It is noticed that the hysteresis phenomenon represents a clue in the search for physical processes responsible for linking the solar activity to the near-Earth and geomagnetic responses.

## 1. Introduction

The Sun and geomagnetic activity are related through the solar wind, a fully ionized magnetized plasma which travels from the Sun to the Earth in a few days. It is well known that when a coronal mass ejection (CME) erupts from the Sun, or a high-speed solar wind stream is emitted and travels in the interplanetary space, they may encounter the Earth, interact with its magnetosphere, and give rise to geomagnetic activity (disturbances of the Earth's magnetic field). Geomagnetic indices, such as $A_p$ (Bartels *et al*., 1939), *aa* (Mayaud, 1972), and $D_{st}$ (Sugiura, 1964), to name a few, are measures of this geomagnetic activity occurring over short periods of time (Mayaud, 1980).

Geomagnetic activity reflects conditions in the solar wind which in turn are influenced by both long-term changes and transient activities on the Sun *(e.g.*, Feynman and Crooker, 1978; Richardson *et al.,* 2000; Svalgaard and Cliver, 2010; Richardson and Cane, 2012). The study of the statistical properties of solar activity and their relations to those in the geomagnetic indices has attracted growing interest (*e.g*., Stamper *et al*., 1999; Kovacs *et al*., 2001; Lui, 2002; Obridko and Shelting, 2009; Kirov *et al*., 2013). There are many studies dealing with trends in geomagnetic indices (Vennerstrom, 2000; Cliver *et al*., 2002; Rouillard *et al.,* 2007; Demetrescu and Dobrica, 2008; among others) but not so many dealing with trends in amplitude and phase of the known oscillations in solar and geomagnetic activity parameters.

Variations in solar activity are traced by measuring sunspot numbers and areas (Hoyt and Schatten, 1998; Hathaway, 2002; Sarychev and Roshchina, 2006; Li, *et al*., 2009; Zieba and Nieckarz, 2014), solar flare index (Singh, Singh, and Badruddin, 2008; Kilcik *et al*., 2010; Yan *et al*., 2012), CMEs (Tousey, 1973; Munro *et al.*, 1979; Webb and Howard, 1994; Cliver and Hudson, 2002; Kane, 2006), 10.7 cm solar radio flux (Wintoft, 2011; Deng *et al*., 2013), total and spectral solar irradiance (Lean *et al.,* 1995; Wenzler, Solanki, and Krivova, 2009; Ball *et al*., 2012), *etc*. All these solar and geomagnetic activity indices display correlative relationships with one another.

Kilcik *et al*. (2011) introduced the "Maximum CME Speed Index" (MCMESI) as a new solar activity indicator closely correlated with the solar and geomagnetic activities. In this study, we explore the relation between MCMESI and some solar, solar wind, and geomagnetic activity indices, namely the international sunspot number, total sunspot area, sunspot magnetic field, and flare index (solar indices), the solar wind speed and interplanetary magnetic field strength (interplanetary indices), the geomagnetic $A_p$ index and disturbance storm time $D_{st}$ index (geomagnetic indices), during solar cycle 23 and the ascending part of cycle 24. Comparisons between these indices and MCMESI indicate significant temporal variations on solar-cycle time scales. As a result, dissimilar patterns of such indices cause hysteresis (loop like structure) dependences between them.

Although the shape of hysteresis curves among several indices has been extensively studied in the past (Bachmann and White, 1994; Özgüç and Ataç, 2003; Bachmann *et al.,* 2004; Kane, 2003, 2011; Suyal *et al*., 2012; Özgüç, Kilcik, and Rozelot, 2012; Ramesh and Vasantharaju, 2014), this is the first time that such relations are seen between the solar activity index MCMESI and some solar, geomagnetic, and interplanetary indices. We recall the definitions of the standard indices which we use in this study and of the one which has been more recently introduced, in Section 2 of this paper. In Section 3, we analyze the MCMESI and geomagnetic and interplanetary space indices. In Section 4, the hysteresis pattern between MCMESI and the other indices is explored. The discussion and conclusions are given in Sections 5 and 6.

**2. Data**
Our study covers the time period since 1996 when data from the *Large Angle and Spectrometric Coronagraph* (LASCO; Brueckner *et al.*, 1995) on the *Solar and Heliospheric Observatory* (SOHO) mission are available. This interval includes the whole of solar cycle 23, the ascending branch of the current cycle 24, and of course the long minimum which occurred between cycles 23 and 24. For all indices we use their yearly averages, to avoid unnecessary details and eventual seasonal variations. The three groups of indices which are used in this study can be briefly described as follows:

2.1 Solar Indices

i) Maximum CME Speed Index (MCMESI) – our main index which we compare to all other indices. The MCMESI data are derived as compiled in the SOHO LASCO CME catalog (Yashiro *et al.*, 2004). The determination of the MCMESI is based on the measurements of the highest daily linear CME speed averaged over one month (for more details, see Kilcik *et al.*, 2011), and then yearly averaged from the monthly averages.

ii) The international sunspot number from the World Data Center for the production, preservation, and dissemination of the international sunspot number (http://sidc.oma.be/silso/), and the sunspot area from the combined Royal Observatory Greenwich - USAF/NOAA sunspot data base (http://solarscience.msfc.nasa.gov/greenwch/sunspot_area.txt) are used for reference to the sunspot cycle phases.

iii) Sunspots are seats of strong magnetic fields. Most solar flares and coronal mass ejections originate from magnetically active regions around sunspots. Here we use the data set of magnetic fields in sunspot umbrae (MF) compiled from measurements in seven observatories in the former Soviet Union. Similar to the determination of MCMESI, the magnetic field of the sunspot umbra with the strongest field for each day of observations is selected. This maximum daily field strength is averaged over one month and then the monthly averages are averaged over one year (Pevtsov *et al.*, 2011).

iv) The solar flare index FI (http://www.ngdc.noaa.gov/stp/space-weather/solar-data/solar-features/solar-flares/index/flare-index/) was introduced by Kleczek (1952) to quantify the daily flare activity over a 24-h period, and represents the intensity scale of the importance of a flare in Hα and the duration of the flare in minutes (for more details, see Özgüç, Ataç, and Rybak, 2002).

2.2 Interplanetary Indices
v) The parameters of the solar wind have been measured *in situ* by a number of Earth orbiting satellites since the sixties of the 20th century. In this study we use the solar wind speed $V_{sw}$ and magnetic field strength $B$ taken from http://omniweb.gsfc.nasa.gov/form/dx1.html, yearly averages of the daily values.

2.3 Geomagnetic Indices
vi) The geomagnetic $A_p$-index is a measure of the level of geomagnetic activity over the globe. Here we use the yearly $A_p$ values averaged over the daily $A_p$ values for the whole year. The daily $A_p$-value is obtained by averaging the eight 3-h values of $a_p$ for each day. In turn, $a_p$ is the linear analogue of the global $K_p$ index which is obtained as the mean value of the disturbance levels in the two horizontal field components, observed at 13 selected subauroral stations (Siebert, 1971). The data series of monthly and yearly $A_p$ values used here is available at ftp://ftp.gfz-potsdam.de/pub/home/obs/kp-ap/ap_monyr.ave/.

vii) The disturbance storm time $D_{st}$ index is derived from a network of near-equatorial geomagnetic observatories and measures the intensity of the globally symmetrical equatorial ring current. Its hourly values have been continuously computed at the World Data Center WDC-C2 at Kyoto, Japan, since the International Geophysical Year 1957. We use the yearly averages of the daily values of $D_{st}$ as available at http://omniweb.gsfc.nasa.gov/form/ dx1.html.

## 3. Analysis

Based on SOHO/LASCO observations from 1996 to 2001, Gopalswamy *et al.* (2004) found that both the number of CMEs and their speeds increase from sunspot minimum to sunspot maximum. Figure 1 presents the temporal variations of (a) the CME parameters, (b) solar activity indicators, (c) solar wind parameters, and (d) geomagnetic activity indices, together with the number and total area of sunspots. For the sake of comparison, all indices in Figures 1a-d are presented in units of standard deviations from their mean values.

It should be noted that here only CMEs wider than $30^o$ are counted to avoid instrumental effects. As LASCO remained the only instrument operating aboard SOHO during the late descending and minimum phase of cycle 23, its image cadence has been improved to ≈12 minutes, compared to 20-30 min in 1996. As a result, fainter/smaller CMEs are registered in the recent years. However if only CMEs wider than $30^o$ are considered, the real solar cycle variation can be seen (N. Gopalswamy, private communication, 2013).

The following features may be noted:
- The popular 11-year Schwabe solar cycle is more or less clearly seen in all series plotted in Figure 1, but despite the similarity among the considered indices, their peak occurrences do not coincide well.
- The CME number and MCMESI lag behind the sunspot number maximum but coincide with the sunspot area maximum (Figure 1a). Actually, as shown by Gnevyshev (1963) and Antalova and Gnevyshev (1965), and explained by Georgieva (2011), all sunspot cycles are double peaked, but in some of them the peaks are too close to be distinguished when the sunspots are averaged over all latitudes. In cycle 23 both the sunspot number and the sunspot area display well expressed double maxima (see e.g. Figure 1 in Ramesh, 2010), which are however smeared in yearly averages and not evident in our Figure 1. The sunspot number's first maximum in 2000 is higher than the second one, while the opposite is true for the total sunspot area. The CME parameters better correlate with the total sunspot area.

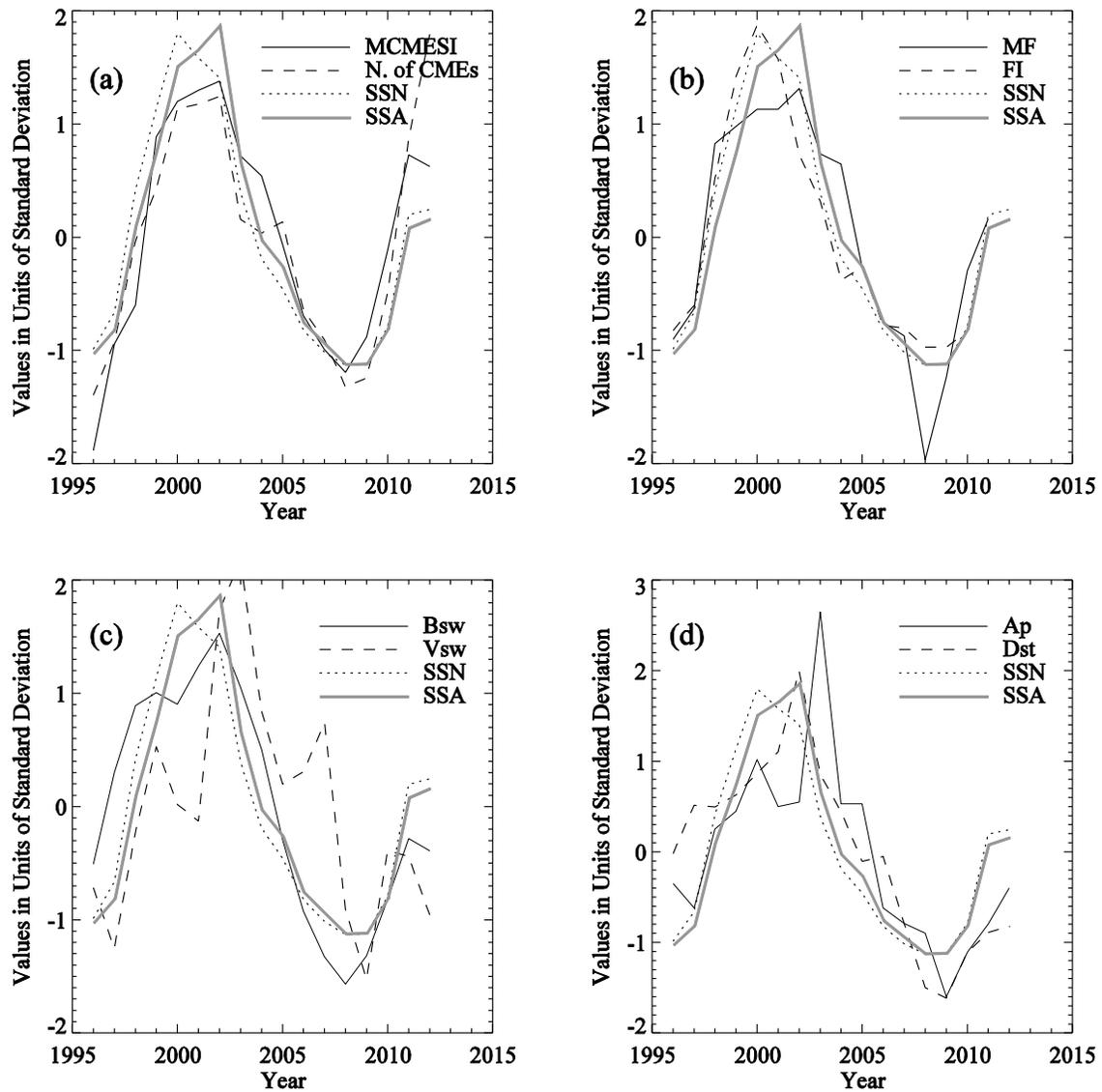

*Figure 1. Time evolution of MCMESI (solid line), sunspot number (dotted line) and area (thick solid line), together with: (a) yearly number of CMEs (red diamonds); (b) sunspot magnetic field MF (solid line) and Flare index FI (dashed line); (c) solar wind magnetic field B (solid line) and speed (dashed line); (d) geomagnetic Ap-index (solid line) and Dst-index (dashed line). All indices are presented in units of standard deviations from the respective mean values.*

- FI is the only index which has a maximum coinciding with the maximum of sunspot number during cycle 23 (Figure1b).

- The sunspot magnetic field MF (Figure 1b), the solar wind magnetic field $B$ (Figure 1c), as well as (the absolute value of) the geomagnetic $D_{st}$ index (Figure 1d) have maxima coinciding with the maximum of the total sunspot area.

- The year of the sunspot number minimum of solar cycle 23/24 (2008) is the year of the minima of all studied solar indices and all interplanetary indices except $V_{sw}$ (to be discussed later). On the other hand, neither of the two considered geomagnetic indices ($A_p$ and $D_{st}$) coincided with the sunspot number minimum; they have their minima a year later.

- The evolutionary pattern of $V_{sw}$ (Figure 1c) seems to differ from all studied solar and interplanetary indices with large variations in both the ascending and descending phases of cycle 23. This pattern continues also during the ascending part of cycle 24. Year 2003 shows the highest value of the averaged solar wind speed which is suggested to be related to several coronal holes which are known to generate recurrent high-speed wind streams (Echer *et al*., 2004). The geomagnetic $A_p$ index (Figure 1d) seems to show some similarity to the solar wind speed $V_{sw}$ in its main maximum in 2003, in the earlier maximum in 1999, and in the minimum in 1997, but unlike $V_{sw}$, it does not display maxima in 2007 and 2010, nor a minimum in 2012.

The above figures support the notion of Ramesh and Rohini (2008) that the sunspot areas better represent the solar cycle than the sunspot number, because in the sunspot number equal weights are given to spots of all sizes. Indeed, when the sunspot areas are used as an indicator of sunspot activity, the CME activity is delayed just 2 to 3 months after the sunspot activity (Ramesh, 2010), as compared to 6-12 months if the sunspot numbers are used. (Raychaudhuri, 2005; Robbrecht *et al*., 2009).

- The relation between MCMESI and the solar, interplanetary, and geomagnetic indices is non-linear. For example, higher relative values of MCMESI correspond to lower sunspot number, sunspot magnetic field, and flare index during the ascending phase of sunspot cycle 24 than during the ascending phase of sunspot cycle 23, and to equal relative values of MF and lower values of sunspot number and flare index during the descending phase of sunspot cycle 23. It can be therefore expected that the relationship between MCMESI and other solar, interplanetary, and geomagnetic indices may display hysteresis-like patterns.

**4. Hysteresis Pattern**

Any two indices with similar or proportionate rise and fall would result in a straight line in their scatter plot. Instead, they depict a loop-like pattern if there is a temporal offset in their overall rise and fall (Figure 1 of Bachman *et al*., 2004). The separation between the two branches of the loop is proportional to the time lag between the two indices (the longer the lag, the wider the loop), and the direction of the loop rotation with evolving time indicates which of the indices lags behind the other one. If the rotation is clockwise, the variable plotted along the *x*-axis lags behind the variable plotted along the *y*-axis, and if the rotation is counter-clockwise, the variable along the *y*-axis lags behind the variable along the *x*-axis.

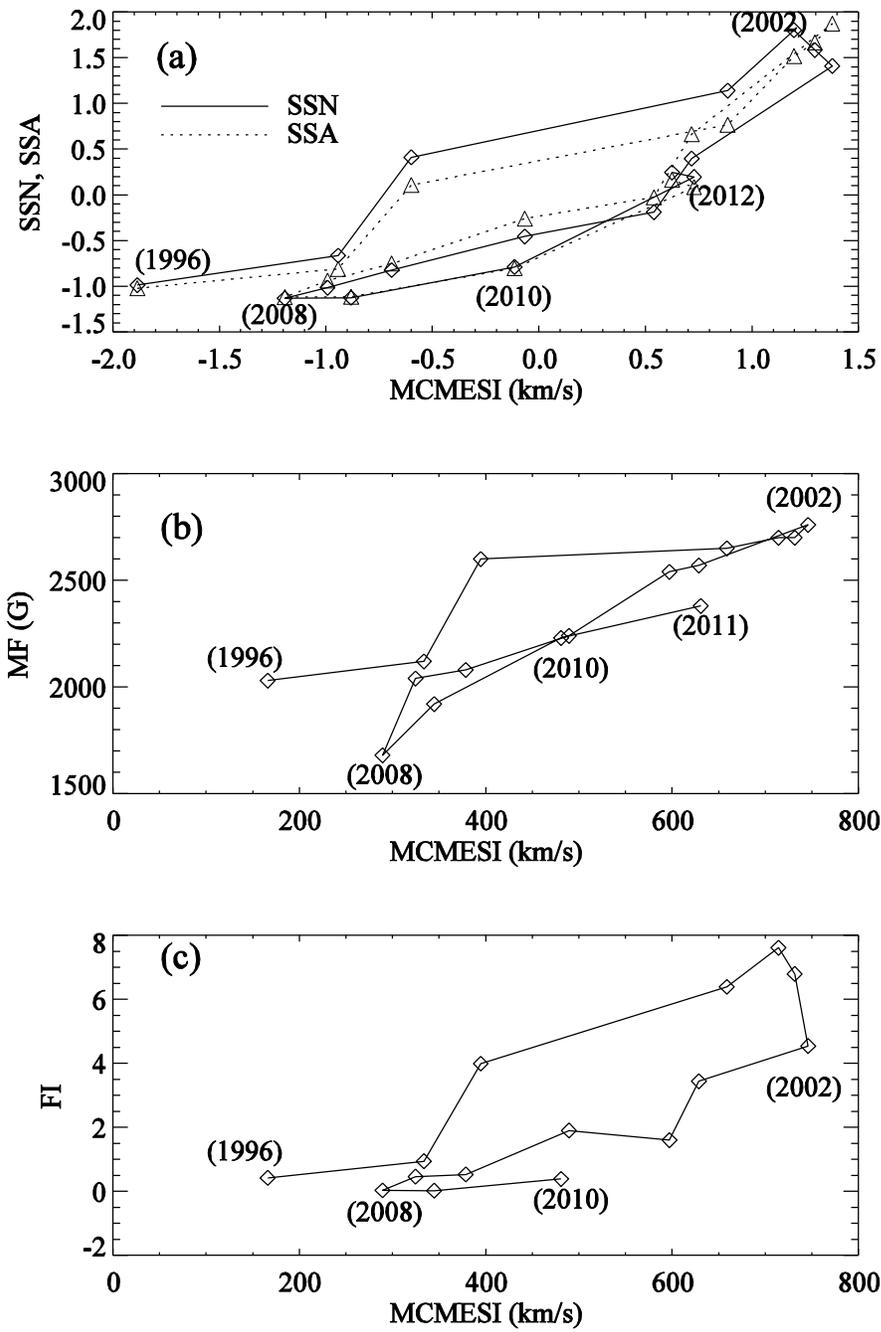

*Figure 2. Scatter plots showing the hysteresis phenomenon for MCMESI versus (a) sunspot number (solid line) and area (dotted line), (b) sunspot magnetic field MF, (c) flare index FI.*

4.1. Solar Parameters

Here we investigate the hysteresis-like patterns to test the nature of relationship between MCMESI (chosen to be the abscissa of all plots with identical scale in Figures 2-4) and solar, interplanetary, and geomagnetic parameters. Figure 2 presents the relationship between MCMESI and the sunspot number and area (Figure 2a), the sunspot magnetic field MF (Figure 2b), and the flare index FI (Figure 2c).

The plotted curves in Figures 2-4 clearly exhibit loop patterns. The time evolution of yearly MCMESI does not follow the same path in the ascending and the descending parts of the solar cycles when it is plotted against the other indices. Arrows indicate the direction of time. Several features can be noted in Figures 2a-c.

The hysteresis loop is clockwise when comparing MCMESI with the sunspot number and area (Figure 2a), which means that MCMESI lags behind sunspot number and area, in accordance with the model of Wheatland and Litvinenko (2001). The loop is wider for the sunspot number (solid line) than for the sunspot area (dotted line), confirming that the lag of the CME activity behind the sunspot area is less than the lag behind the sunspot number (Ramesh, 2010).

The hysteresis pattern between MCMESI and the sunspot magnetic field (Figure 2b) is similar to the ones between MCMESI and the sunspot number and between MCMESI and the total sunspot area (Figure 2a). This is to be expected because the logarithm of sunspot area is linearly correlated to its magnetic field (Pevtsov *et al.*, 2014), and the total area of sunspots is correlated to the number of sunspots (Hathaway, 2010). Both for the sunspot area and MF, the width of the loop is not constant. In the years around the solar maximum (1999-2003) there is practically no hysteresis, which can be interpreted as no or very small delay between the sunspot MF and the CME activity.

An interesting problem is the hysteresis between MCMESI and the flare index FI (Figure 2c). Though both solar flares and CMEs are processes releasing free energy stored in the corona, their temporal variations seem to differ, with the CME maximum speed lagging behind the FI by two years (Figures 1b and 2c). We will come back to this in Section 5.

4.2. Solar Wind Parameters

Figure 3 demonstrates the hysteresis patterns between MCMESI and solar wind parameters (magnetic field and speed). As in the case of the solar activity indices, MCMESI lags behind the magnitude of the interplanetary magnetic field which is a continuation of the large scale solar magnetic field (Figure 3a). Around the sunspot maximum (1999-2003), the direction of the loop changes. The direction of circulation of the correlation between MCMESI and $V_{sw}$ follows a path that is opposite to those of the three solar activity indicators and the interplanetary magnetic field, *e.g.*, the ascending path follows a lower track than for the descending one (counterclockwise rotation). This means that the variations in MCMESI precede the variations in the solar wind speed. Moreover, the width of the loop is not constant. It is biggest in 2003 when the highest solar wind speeds were recorded.

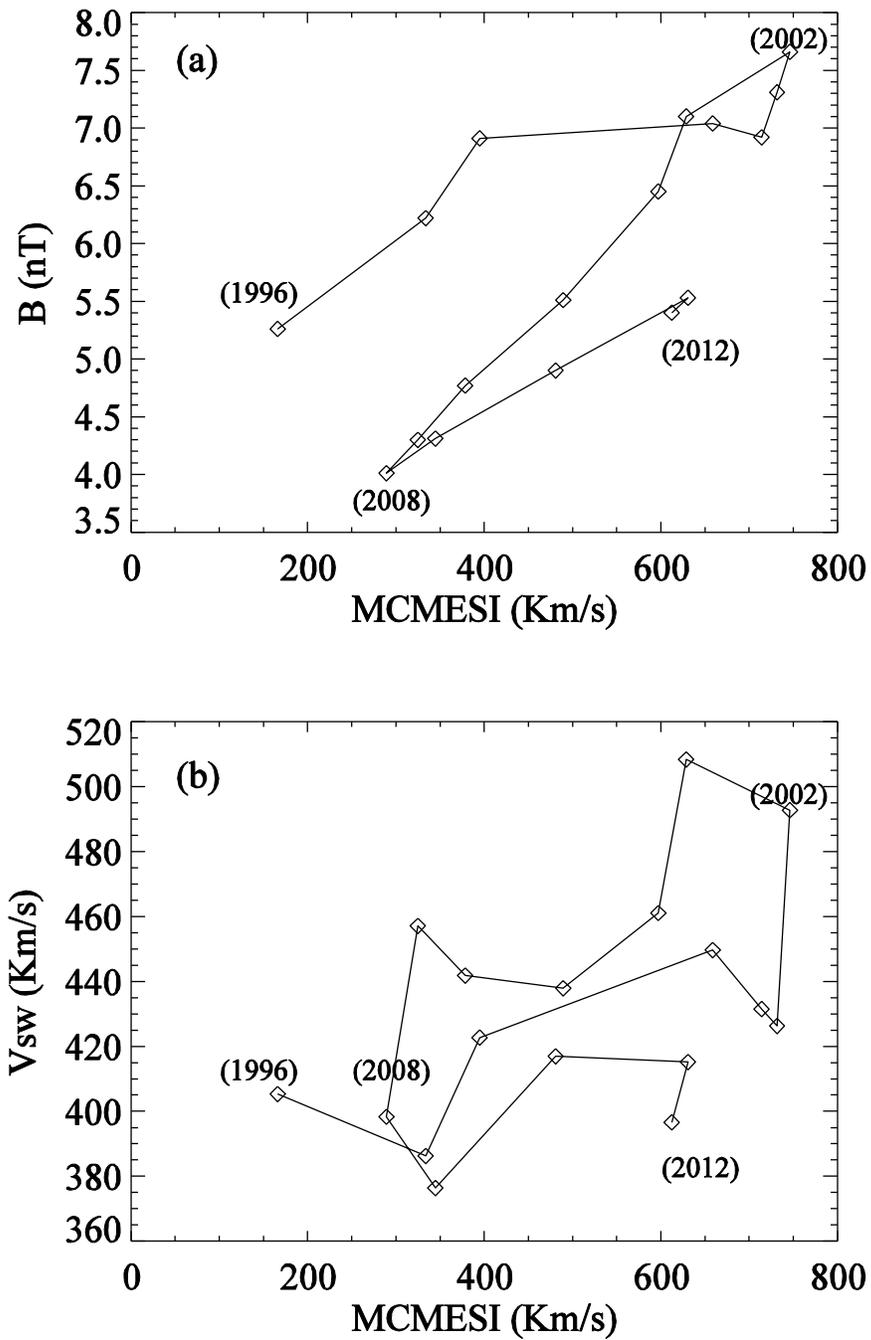

*Figure 3.* Scatter plots showing the hysteresis phenomenon for MCMESI versus (a) magnitude of the interplanetary magnetic field scalar B, (b) solar wind velocity Vsw.

4.3. Geomagnetic Indices

The hysteresis pattern of MCMESI *versus* the geomagnetic $D_{st}$-index (Figure 4a) is very similar to the one with $B$ (compare with Figure 3a) but with some differences in width. Namely, $B$ depicts a broad loop while $D_{st}$ depicts narrow hysteresis loops. The direction of rotation during cycle 23 is clockwise and changes to counter-clockwise around the sunspot maximum. The change in the direction of the loop rotation in the same period is also seen in the scatter plot of MCMESI *versus* $a_p$-index. As in the case of $V_{sw}$, the separation is wider around the maximum of solar wind speed (Figure 4b). However, the basic rotation is in the same direction as for all other indices except $V_{sw}$.

We cannot evaluate the hysteresis patterns in cycle 24 because a hysteresis loop can be visualized only when both the ascending and descending phases of the cycle data are available, and the descending phase of cycle 24 has just begun. But we have plotted the points until 2012 for comparison, and we find that the relations between the solar indices in cycle 24 are different from the ones in cycle 23. During the ascending branch of cycle 24, all the indices that we studied generally show low activity compared to cycle 23. In all cases, the points are not placed on the same line but with lower values and follow a different line, and the same MCMESI values in cycle 24 correspond to much smaller sunspot, solar wind, and geomagnetic activity parameters.

## 5. Discussion

This study started with an aim of close examination of the relationship between solar, interplanetary, and geomagnetic indices with a solar activity parameter (here the MCMESI) for the period of 17 years over solar cycle 23, the long minimum which followed it, and the ascending branch of cycle 24. We use three sets of indices to correlate with MCMESI: the international sunspot number, the total sunspot area, the sunspot magnetic field, and the flare index (solar activity indices); the interplanetary magnetic field magnitude and the solar wind speed (interplanetary indices); and the disturbance time index $D_{st}$ and $A_p$–index (geomagnetic indices). It is evident from simple comparisons of the respective mean yearly data series that the MCMESI is to the greatest extent correlated with the sunspot magnetic field MF and total sunspot area, with the magnitude of the interplanetary magnetic field $B$, and with the geomagnetic $D_{st}$ index; to a lesser extent with the sunspot number and with the flare index FI; and to a still lesser extent with the solar wind speed $V_{sw}$ and geomagnetic $A_p$ index (Figures 1 a-d). Among the indices here studied, we find hysteresis effects which are approximately regarded as simple phase shifts, and we quantify these phase shifts in terms of lag times behind the leading index, MCMESI.

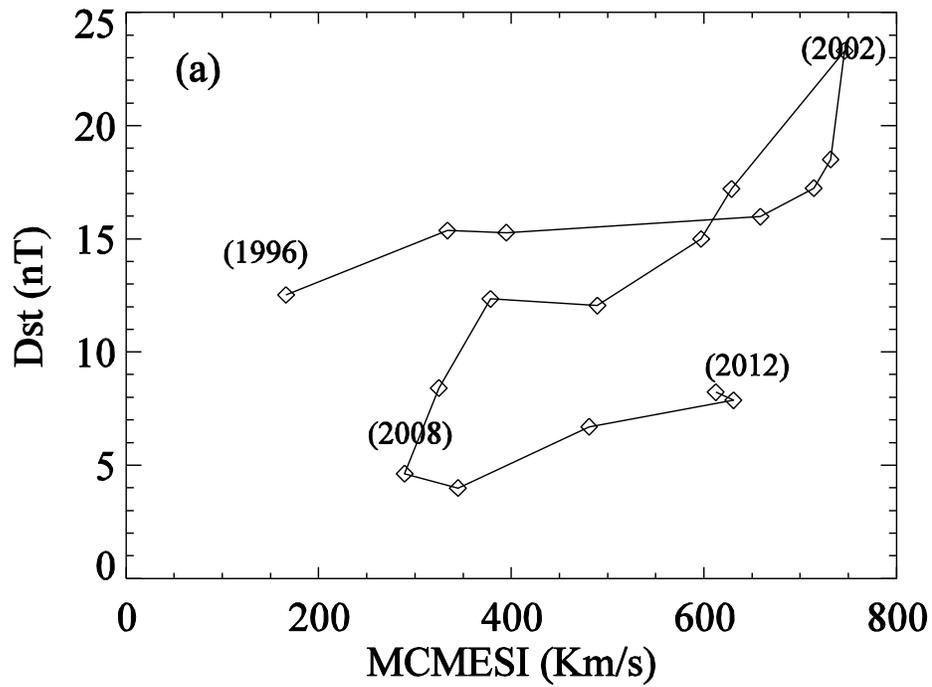

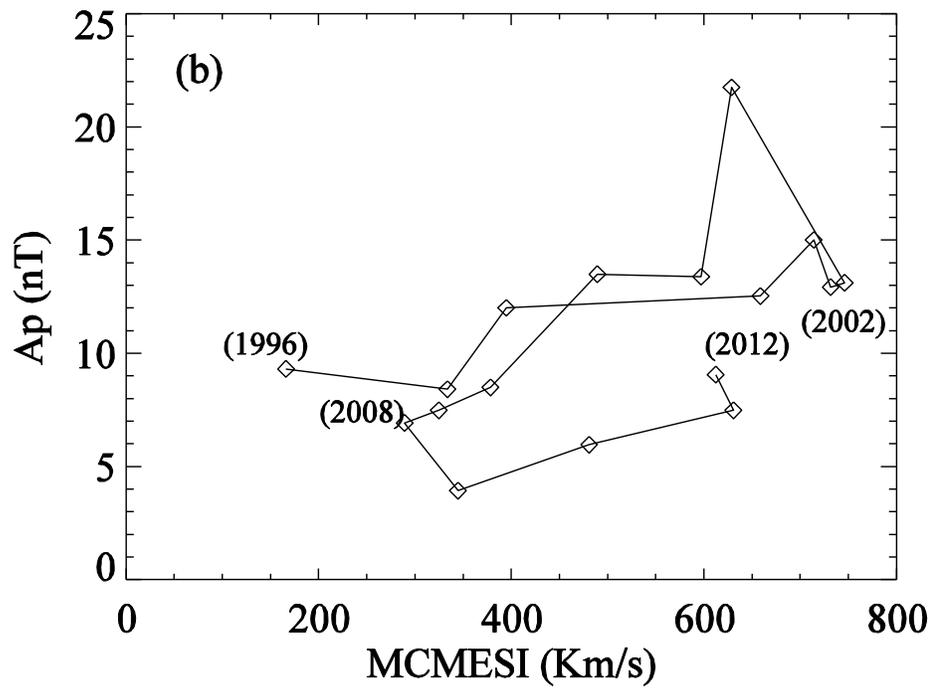

*Figure 4.* Scatter plots showing the hysteresis phenomenon for MCMESI versus geomagnetic activity indices: (a) the (absolute value of) the disturbance time index Dst, (b) Ap-index.

5.1. MCMESI/Solar Activity Indices

The hysteresis between two parameters is generally assumed to represent cause and affect relationships. What can be the reason for the hysteresis in the case of the Sun, if all solar activity indices are manifestations of processes ultimately driven by the action of a common magnetic dynamo mechanism?

It is known that different indices reflect processes occurring at different depths in the Sun. The time for some processes to reach the altitudes at which other processes occur, determines the width of the hysteresis loops between these two processes. It is therefore interesting to estimate why the hysteresis loops between MCMWSI and sunspot number, and between MCMESI and total sunspot area are different, if the sunspot number and the sunspot area both characterize sunspots. The answer was provided as early as 1974 by Prokakis (1974) who showed that sunspots with different sizes have their source regions at different depths. The size distribution (and therefore the depth distribution) of sunspots is reflected in sunspot total areas but not in the sunspot number which gives equal weights to sunspots of different sizes.

While sunspots are surface or photospheric phenomena, flares and CMEs originate above the photosphere, in the corona (Kane, 2006). For the flares and CMEs to occur, there should be enough free energy stored in the corona. Assuming that the sunspot number is a proxy for the energy supply to the corona, and solar flares and CMEs are the dominant mechanism by which the free energy stored in the corona is released, Wheatland and Litvinenko (2001) developed a model of the energy balance in the solar corona over the sunspot cycle which predicts that the free energy in the corona should lag behind the variations in the energy supply to the corona. The time needed for the energy supplied to the corona to reach the high enough level to produce flares and CMEs is the time lag between the sunspot numbers/areas and the parameters characterizing flares/CMEs, and can account for the observed hysteresis effects between MCMESI and sunspot number and area.

The time needed for the storage of enough free energy in the corona to produce CMEs explains also the hysteresis effects between MCMESI and sunspot magnetic fields. Both sunspots and most of the CME source regions are embedded in solar active regions (ARs) – sites of strong and complex magnetic fields. The sunspot magnetic fields are indicative of the magnetic fields of ARs, and the speed of a CME depends on the magnetic field of its source region, with higher speed CMEs originating from source regions with stronger magnetic fields (Gopalswamy *et al.*, 2008). The sunspot magnetic fields may be an even better indicator of the energy supplied to the corona than the sunspot number. Figure 2b demonstrates that, around the sunspot maximum, the sunspot/AR magnetic fields and CME speeds are immediately (in yearly averages) correlated, with very small delay. This change in the delay in the course of the sunspot cycle may mean that energy is very quickly transmitted from the photosphere into the corona around sunspot maximum, and much more slowly around sunspot minimum, evidenced by the increased width of the hysteresis loop.

The correlation between the AR magnetic field and the CME speed provides an explanation of another interesting fact. It is well known that the number and speed of CMEs follow the sunspot cycle (Gopalswamy *et al*., 2004). However, Gao, Li, and Zhang (2014) showed that though the maximum sunspot number of cycle 24 is significantly smaller than that of cycle 23, the number of CMEs in cycle 24 is larger than that in cycle 23, while the speeds of CMEs in cycle 24 are smaller than those in cycle 23. The explanation is based on the fact that strong magnetic fields in CME source regions are constraints for the CME initiation, because a CME must first overcome the background magnetic field before it can erupt. With weak magnetic field of ARs, the constraint on the background magnetic fields for the eruptive events is also weak, which makes it easier for the CMEs to escape outward, and leads to a bigger number of CMEs (Gao, Li, and Zhang, 2014). On the other hand, as pointed out above, the speed of a CME depends on the magnetic field of its source region. Therefore, weaker AR magnetic fields in cycle 24 compared to cycle 23, if they are weaker indeed, could explain both the increased number of CMEs, and the lower CME speeds.

Are the AR magnetic fields weaker in cycle 24 as compared to cycle 23? The question of sunspot cycle and secular variations of sunspot magnetic fields has been extensively studied during the last decade. Kiess, Rezaei, and Schmidt (2014) used data of the *Helioseismic and Magnetic Imager* onboard the *Solar Dynamics Observatory* and selected all sunspots between May 2010 and October 2012, using one image per day; they reported that, from the investigation of umbral area, magnetic field, magnetic flux, and umbral intensity of the sunspots of the rising phase of cycle 24, they do not find a significant difference to the previous cycle, and hence no indication for a long-term decline of solar activity. However, Livingston and Penn (2009) and Penn and Livingston (2011), using the Zeeman-split 1564.8 nm Fe I spectral line at the NSO Kitt Peak McMath-Pierce telescope, found that the magnetic field strength in sunspots has been decreasing in time since 1990's, with no dependence on the solar cycle. Moreover, these authors predicted that if this decrease continues at its current rate, the sunspot magnetic field will soon fall under the limit of 1500 gauss below which no sunspots are observed. On the other hand, Pevtsov *et al*. (2011) employed historic synoptic data sets from seven observatories in the former USSR covering the period from 1957 to 2011, and found that the sunspot field strengths vary cyclically reaching maxima around sunspot maxima and minima around sunspot minima, with no indication of a secular trend in the last five sunspot maxima (cycles 19-23), but with a well expressed trend in the sunspot minima (Kirov *et al.*, 2013). The explanation of this contradiction was offered by Nagovitsyn *et al*. (2012): While Pevtsov *et al*. (2011) used only the biggest sunspots for their analysis, Livingston and Penn (2009) and Pen and Livingstone (2010) used all visible sunspots. During the period of 1998–2011, the number of large sunspots whose magnetic field does show sunspot cycle variations but no long-term trend, gradually decreased, while the number of small sunspots which have weaker magnetic fields decreasing in time but without sunspot cycle dependence, steadily increased. The observations of Kilcik *et al*. (2014) that there is a very significant decrease in the number of large sunspot groups and in the sunspot counts in large sunspot groups in cycle 24 compared to cycles 21-23 also confirm this result. Taking into account that the sunspot magnetic field is related to the sunspot area (Houtgast and van Sluiyers, 1948), and that the CME speed is higher for higher magnetic field of the AR/CME source region related to sunspot magnetic field (Gopalswamy *et al.*, 2008), it is natural to expect that the decreased sunspot number and sunspot group area (the decreased AR magnetic field) during cycle 24 will lead to an increased number (lower speed) of CMEs, respectively.

Both CMEs and solar flares are solar eruptive phenomena, associated with large-scale, closed magnetic field structures in the corona. These structures change throughout the solar activity cycle following the evolution of the general solar magnetic field. CMEs differ from flares physically, in that the flare represents the luminous output of the process, whereas the CME represents certain ejecta. For a CME to occur, the coronal magnetic field blows open and stretches into a more energetic state, linking temporarily a new part of the solar atmosphere to the solar wind (Hudson and Li, 2010). CMEs are often accompanied by solar flares, and it is believed that if they are not, the lack of association is due to either the flare being behind the solar disc, or the flare being too weak to be registered (Chen (2011) and the references therein). On the other hand, many flares are not associated with CMEs.

Wang and Zhang (2007) used the potential field source-surface model to infer the coronal magnetic field above the source active regions and calculated the flux ratio of low to high corona. They found that the flares without CMEs ("confined events") have a lower ratio than the flares with CMEs ("eruptive events"). Thus, the excess energy can be stored in the corona for subsequent CMEs, and therefore the variations of the CME speeds and energy will be delayed about two years after the flare activity. The hysteresis observed in cycle 23 between MCMESI and FI (Figure 2c) confirms this hypothesis.

5.2. MCMESI/Interplanetary Indices

The solar wind is a continuous outflow of plasma from the Sun due to the lack of hydrostatic equilibrium in the solar corona, leading to its expansion (Parker, 1958). Due to its high conductivity, the solar wind drags the solar magnetic field which is "frozen in" to the plasma. It is the speed of this solar wind and the magnitude of its frozen-in magnetic field which we measure at the Earth's orbit.

Figure 3a demonstrates that during most of cycle 23, the variations of MCMESI lag after the variations of the interplanetary magnetic field magnitude, except for the period around the sunspot maximum when the IMF lags after the MF. It should be reminded here that the solar magnetic field has two components – poloidal and toroidal – which transform into each other, much like potential and kinetic energy in a harmonic oscillator (Parker, 1955). These two components (the one related to the solar open flux reaching the Earth's orbit, and the other to the closed magnetic field configurations giving rise to CMEs) become out of phase in the course of the sunspot cycle. The solar toroidal field whose manifestations are the CMEs, is produced from the solar poloidal field; therefore during most of the sunspot cycle, the CME number and speed are delayed with respect to the solar wind magnetic field, as confirmed by Figure 3a. In turn, the solar poloidal field is produced by the solar toroidal field shortly after the sunspot maximum, therefore the reversed dependence around sunspot maximum.

The solar wind speed is the only index among the ones studied, which lags with respect to MCMESI. The maximum in the solar wind speed occurs during the solar cycle declining phase when large low-latitude coronal holes and equatorward extensions of polar coronal holes emanate fast solar wind close to the ecliptic plane (Wang and Sheeley, 1990), while MCMESI has a maximum during the sunspot maximum phase when the AR magnetic fields (determining the maximum CME speed) are strongest.

5.3. MCMESI/Geomagnetic Indices

The variations observed in the interplanetary space and in the Earth's environment are attributed to variations in solar activity. It is well known since the availability of geomagnetic activity indices (1868 to nowadays) explained by Mayaud (1972, 1980) that one of the best signatures of the solar variability recorded on Earth is geomagnetic activity. On average, the distribution of magnetic disturbances over the solar activity cycles exhibits two peaks, one at the maximum of the sunspot number (or two in sunspot maximum if the sunspot maximum is double-peaked), and another during the descending phase of each cycle. The first peak is due to CMEs whose number is highest during sunspot maximum. The later peak, during the sunspot declining phase, is caused by recurrent high-speed streams from coronal holes (Gonzalez *et al.*, 2002; Echer *et al.*, 2004; Abramenko *et al.*, 2010). In our case, cycle 23 supports this (compare Figures 1a and 1b).

When a magnetic storm occurs, following a CME or a high-speed solar wind stream, one observes several superimposed phenomena. The classical geomagnetic indices $K_p$ (and the similar $A_p$, *aa, etc*.) and $D_{st}$ have been devised to capture magnetic variations of different origins recorded in observatories. These indices are intended to monitor different processes: the position of the auroral oval and (mostly) the intensity of the magnetospheric ring current, respectively, and could therefore be expected to display different time variations (Le Mouël *et al*., 2012). In particular, $A_p$ and similar indices ($K$, $K_p$, *aa.*, *etc*.) respond to both CMEs and high-speed solar wind streams, while the $D_{st}$ index is more sensitive to CMEs (Borovsky and Denton, 2006). Figure 1d confirms this; the $D_{st}$ index has a single maximum in 2002 coinciding with the maximum in sunspot/AR magnetic fields, interplanetary magnetic field magnitude, and MCMESI, while $A_p$ has two maxima coinciding with the maxima in solar wind speed (high-speed solar wind streams) in 2000 and in 2003, in which, however, the number and speed of CMEs and $D_{st}$ had no maxima.

As a result, the hysteresis patterns of MCMESI with $D_{st}$ and $A_p$ geomagnetic indices basically repeat the patterns of MCMESI with solar and interplanetary parameters. The hysteresis pattern of MCMESI *versus* the geomagnetic $D_{st}$-index (Figure 4a) is very similar to the one with *B* (compare with Figure 3a) but with some differences in width. Namely, *B* depicts a broad loop while $D_{st}$ depicts a narrow hysteresis loops. The direction of rotation during cycle 23 is clockwise and changes to counter-clockwise around sunspot maximum. The change inthe direction of the loop rotation in the same period is also seen in the scatter plot of MCMESI *versus* $A_p$-index, and as in $V_{sw}$ the separation is wider around the maximum of solar wind speed. However, the basic rotation is in the same direction as for all other indices except $V_{sw}$.

## 6. Conclusions

The hysteresis phenomenon between a pair of indices, which is due to a temporal offset between their variations, can be used as a diagnostic tool to better understand the chain of phenomena linking various manifestations of solar activity, interplanetary medium, and the Earth's magnetic field. In the present paper we have compared one solar activity parameter (MCMESI – the maximum CME speed) with other solar (sunspot number and total area, sunspot magnetic field, and flare index), interplanetary (solar wind magnetic field and speed), and geomagnetic ($A_p$ and $D_{st)}$ parameters. It should be noted that although we have used yearly average values throughout this study, the assertions can be applied to broader data specifications.

The relations between MCMESI and the other solar, interplanetary, and geomagnetic parameters hint at conclusions as follows:

- The hysteresis-like relationship between MCMESI and other solar activity indices reflects the height propagation of solar activity from the photosphere into the corona. The hysteresis-like relationship between MCMESI and the solar flare index supports the idea that solar flares and CMEs are not elements of a common process of generation and ejection of coronal mass, but rather energy is stored in the process of AR emergence during the sunspot ascending phase which can be released in subsequent CMEs during the sunspot descending phase (Ivanov, 2014).
- The hysteresis patterns between the variations of MCMESI and solar wind speed and magnetic field nicely illustrate the mutual transformations of solar poloidal and toroidal fields (responsible for the solar wind open flux and high speed solar wind streams, and CMEs, respectively).
- Finally, the comparison between the MCMESI hysteresis patterns and the solar wind and geomagnetic indices confirms the effects of different solar drivers on the geomagnetic indices that were designed to reflect different kinds of interactions.

As a conclusion, the investigation of hysteresis-like relationships between various solar, interplanetary, and geomagnetic indices can shed light on the processes taking place between the Sun and the Earth, mediated by the solar wind.